\documentclass{iopjournal}
\usepackage{amsmath} 
\usepackage{mathtools}
\usepackage{graphicx}
\usepackage{subfig}
\usepackage{bbold}   
\usepackage{hyperref}
\usepackage[yyyymmdd]{datetime} 
\usepackage{natbib}		
\usepackage{appendix}

\usepackage{xcolor}

\setcitestyle{authoryear}	
\usepackage{braket}
\renewcommand{\articletype}[1]{{\vspace*{-8mm}\noindent \large \sf #1}\\[2ex]}

\begin{document}

\articletype{Paper} 

\title{Continuum Limits of Lazy Open Quantum Walks}

\author{Lara Janiurek$^1$\orcid{0009-0007-8258-6015}, Viv Kendon$^1$\orcid{0000-0002-6551-3056}}

\affil{$^1$Department of Physics, University of Strathclyde, Glasgow, Scotland}

\email{lara.janiurek@strath.ac.uk}

\begin{abstract}
We derive the continuous spacetime limit of the one dimensional lazy discrete time quantum walk, obtaining explicit macroscopic evolution equations for a three state model in the presence of decoherence. While continuum limits of two state quantum walks are well established, an explicit continuous spacetime formulation for the lazy three state walk, particularly including noise, has not previously been constructed. Using an $\mathcal{SU}$(3) representation of a Grover type coin together with a Lindblad formulation of decoherence acting either on the coin or the spatial subspace, we systematically expand the discrete dynamics in both space and time to obtain continuum master equations governing the coarse grained evolution. The resulting generators yield a genuine partial differential equation description of the walk, going beyond purely probabilistic or spectral correspondences. We show that the unitary limit is governed by a Dirac-type $\mathcal{SU}(3)$ Hamiltonian describing ballistic advection of left and right moving modes coupled by local symmetric mixing, with the rest state acting as an additional internal degree of freedom. Coin dephasing selectively damps internal coherences while preserving coherent spatial transport, whereas spatial dephasing suppresses long range spatial interference and rapidly drives the dynamics toward classical behaviour. This continuum framework clarifies how internal symmetry, rest state coupling, and distinct decoherence channels shape large scale transport in lazy open quantum walks, and provides a foundation for future extensions toward multichannel quantum transport models and quantum inspired algorithms.
\end{abstract}

\section{Introduction}
Quantum walks constitute the quantum analogue of classical random walks and form a versatile framework for modelling quantum transport and computation \citep{kendon_quantum_computation,dimolfetta2021}. Unlike their classical counterparts, quantum walks exhibit interference effects that can lead to quadratic speedups in hitting and mixing times, underpinning their use in quantum algorithms for search and graph analysis \citep{Kempe_Shenvi_Whaley,ambainis2004,doi:10.1137/S0097539705447311}. Both continuous time and discrete time formulations have been developed \citep{Farhi_1998,PhysRevA.48.1687}, with discrete time quantum walks (DTQWs) serving as the most direct analogue of lattice based classical walks. 

In the discrete time setting, the walker occupies a position register with an internal coin space, whose rotation governs directional propagation. The variant relevant for this work is the lazy quantum walk, originally introduced by  \cite{Childs_2009} to demonstrate that the continuous time quantum walk arises as a limit of the discrete model. The inclusion of a rest state introduces a non-zero probability of remaining at the current site, modifying the spectrum of the walk operator and enabling a smooth connection between discrete and continuous time dynamics. Unlike the approach of \cite{Childs_2009}, where the lazy walk was introduced to embed continuous time dynamics at the spectral level, our analysis presented here includes dephasing, and derives an explicit continuous spacetime master equation by expanding the discrete Lindblad map in space and time. This yields a genuine coarse grained partial differential equation (PDE) description rather than a spectral correspondence between discrete and continuous models. 

Beyond foundational interest, lazy walks have also been employed in algorithmic contexts, such as Grover’s search, where self loops at vertices act as effective rest states, influencing success probabilities \citep{Wong_2015, Wong_2018,Giri_2019}. Including a rest state extends the walk to a three channel model that more closely resembles discrete velocity schemes in kinetic theory \citep{Succi2018}. The stationary component provides a minimal setting where propagating and stationary dynamics coexist \citep{DiMolfetta2012}. This may therefore serve as a foundation for future developments using open quantum walks to model coarse grained transport or hydrodynamic behaviour. Continuous spacetime limits provide a bridge between discrete quantum walks and continuum physical theories \citep{Manighalam2020, DiMolfetta2019}. By expanding the discrete step operator in powers of a small parameter that controls the time and lattice spacing, one can identify the effective PDE governing large scale dynamics, ranging from Dirac-type to dispersive behaviour, depending on the chosen coin and scaling \citep{Jolly_2023, Dirac_walk_limit}. This approach makes it possible to interpret microscopic unitary rules as discrete simulators of macroscopic behaviour.

Open quantum walks (OQWs) generalise the framework to include environmental noise or engineered dissipation through completely positive trace preserving (CPTP) maps. Environment induced errors in discrete quantum walks were first studied by \cite{Dur_2002} to model errors in proposed experimental implementations.  \cite{Kendon-useful} found that decoherence could improve the algorithmic properties of quantum walks by providing a smoother uniform sampling that was still quadratically faster than classical.  Dissipation has been leveraged more generally for quantum computing and quantum state preparation \citep{Almut_dissipation,Verstraete2009}. More general open system quantum walk models were introduced by \cite{Attal_2012,ATTAL20121545,Sinayskiy_2012}. They provide a natural setting for describing transport in open quantum systems and have found applications in dissipative computing, state transfer, and biological energy transport \citep{Mohseni2008,Sinayskiy2014}. 

Extensions of this framework include lazy open quantum walks, which incorporate a rest state alongside the propagating channels. In particular, \cite{Kemp2020}, constructed a lazy OQW model derived from a Lindblad master equation and proved a central limit theorem demonstrating Gaussian asymptotics with an analytically computable covariance. Such analyses, however, remain microscopic, characterising asymptotic statistics without yielding explicit continuum PDEs for the coarse grained evolution. \cite{Tude_2022b} examine multiple forms of decoherence acting on the three state walk and report how these noise mechanisms change its spreading profile, localisation behaviour, and sensitivity to the rest state. Another closely related line of work is that of \cite{Sinayskiy2015}, who derived a discrete time open quantum walk from a microscopic system-bath model. Both of these analyses stay within the discrete-time framework. 

The inclusion of decoherence or dephasing in DTQWs provides a natural bridge between coherent quantum transport and classical diffusive behaviour. In the present work, we focus on this intermediate regime, where the quantum walk exhibits partial coherence and remains amenable to continuum analysis. Such limits are particularly interesting because they form the conceptual foundation of more complex open and lazy quantum walks, which have been shown to interpolate between ballistic quantum spreading and diffusive or Gaussian transport \citep{Kemp2020}.  Previous studies have already demonstrated that standard quantum walks can admit a hydrodynamic interpretation in the continuum limit, with emergent flow-like or Madelung type equations governing coarse grained dynamics \citep{hatifi2017quantumwalkhydrodynamics}. Building on these ideas, dephased or lazy variants provide a controlled way to introduce dissipation, which is central to hydrodynamic and kinetic descriptions of fluids. Consequently, the simple dephased limit provides a controlled setting for probing the onset of hydrodynamic behaviour in quantum lattice models. While the broader application to quantum or quantum–inspired fluid algorithms
remain open, these results connect naturally to ongoing work on lattice and particle based quantum transport methods \citep{Au_Yeung_2025,
itani2021analysiscarlemannlinearizationlattice}.

In this paper, we analyse a lazy DTQW subject to two distinct noise channels: one acting on the coin degree of freedom and another on the spatial register. By performing a systematic scaling analysis of the discrete evolution, we derive the corresponding continuum master equation governing the coarse grained state. This provides insight into the behaviour of lazy quantum walks with decoherence and complements existing probabilistic approaches, which establish Gaussian limits but not explicit macroscopic dynamics. For completeness, we derive the continuum limit for both the coin and spatial decoherence models. However, previous studies have shown that coin dephasing preserves spatial coherence while suppressing only internal superpositions, making it more useful for modelling controlled decoherence in quantum transport and algorithmic settings \citep{Paz_Lopez_deco,Love2005FromDT,Kendon-useful}. Here we present the continuous spacetime limit of the lazy three state walk in a full $\mathcal{SU}(3)$ representation, where the unitary and dephasing generators emerge naturally from the Gell Mann structure. This offers a clear route to continuum models for multichannel quantum walks and highlights how internal symmetry and coupling shape large scale dynamics.

The paper is organised as follows. Section~\ref{intro_DTQW} reviews the lazy DTQW and introduces the notation used throughout, including the treatment of decoherence. Section~\ref{Section_cont_limit} develops the continuous spacetime limit of the lazy quantum walk.  Section~\ref{discussion-sec} discusses the results and the main conclusions.

\section{Lazy Discrete Time Quantum Walks}\label{intro_DTQW}

We consider a lazy discrete time quantum walk (DTQW) on the line with Hilbert space \(\mathcal{H}_{xc} = \mathcal{H}_x \otimes \mathcal{H}_c\). 
The position space \(\mathcal{H}_x\) is spanned by \(\{|x\rangle : x \in \Delta_x\mathbb{Z}\}\) with lattice spacing \(\Delta_x>0\). The coin Hilbert space is three dimensional, 
\(\mathcal{H}_c = \mathrm{span}\{|L\rangle,|S\rangle,|R\rangle\}\), corresponding to ``right'', ``stay'', and ``left'' moves. 
The global state evolves by
\begin{equation}\label{wavefun_evolution_eqn}
    |\psi(t+1)\rangle = \mathbf{\hat U} \,|\psi(t)\rangle
    = \mathbf{\hat S} \, (\mathbb{I}\otimes \mathbf{\mathbf{\hat{C}}})\,|\psi(t)\rangle 
\end{equation}
where \(\mathbf{\mathbf{\hat{C}}} \in \mathcal{U}(3)\) is the coin operator, \(\mathbf{\hat S}\) is the conditional shift, and \(\mathbf{\hat U}\) is the unitary evolution operator. 
The conditional shift operator moves the particle left, right, or leaves it in place:
\begin{equation}\label{shift-basic}
\mathbf{\hat S} = 
\sum_x \Big(
|x{+}\Delta_x\rangle\langle x| \otimes |R\rangle\langle R| + |x\rangle\langle x| \otimes |S\rangle\langle S|
+ |x{-}\Delta_x\rangle\langle x| \otimes |L\rangle\langle L|
\Big)
\end{equation}
The coin operator is a unitary operator that controls the dynamics of the DTQW. A common choice is the Grover coin, with three-dimensional form
\begin{equation}\label{grover-coin}
\mathbf{\mathbf{\hat{C}}}_G =  2|u\rangle\langle u| -\mathbb{I}_3 =  \frac{1}{3}
\begin{pmatrix}
-1 & 2 & 2 \\
2 & -1 & 2 \\
2 & 2 & -1
\end{pmatrix},
\end{equation}
where 
\begin{equation}\label{u_vector}
    |u\rangle = \tfrac{1}{\sqrt{3}}(1,1,1)^{\mathsf T}.
\end{equation} The coin operator is as unbiased as possible across all three outcomes \citep{Saha2021}. The Grover coin provides a simple and highly symmetric choice for the internal dynamics of the walk. Different coin parametrisations would, in general, lead to distinct effective behaviours in the continuum limit. In the present work, we focus on the symmetric Grover-type coin in order to preserve translational invariance and maintain a balanced coupling between all internal states. This choice ensures that any emergent large scale dynamics arise solely from the interplay between coherent evolution and decoherence, rather than from asymmetries in the underlying coin operation. Nonetheless, the symmetry can be intentionally broken if one wishes to introduce directional bias or model anisotropic effects. 
\begin{figure}[!htb]
    \centering 
    \includegraphics[width=0.6\linewidth]{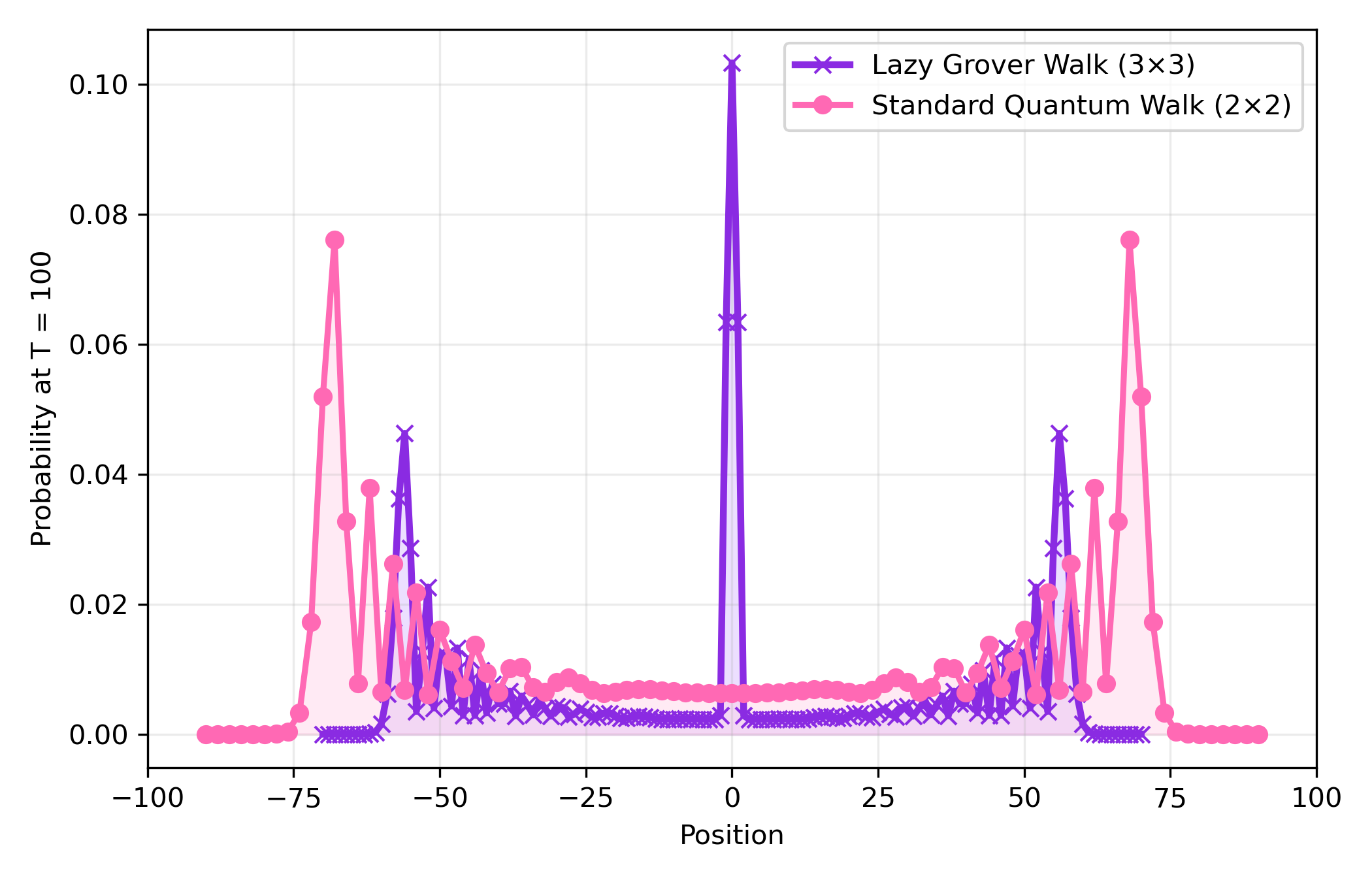}
    \caption{Probability distribution of the three–state lazy DTQW and the standard two–state DTQW after $t=100$ steps. The standard walk uses an  initial state $\ket{0}\otimes\frac{1}{\sqrt{2}}(\ket{L}+i\ket{R})$. The initial state of the lazy walk is an equal magnitude Fourier–symmetric superposition, $|\psi_0\rangle = |0\rangle \otimes \frac{1}{\sqrt{3}}\bigl(|L\rangle + \omega\,|S\rangle + \omega^{2}\,|R\rangle\bigr),$ with $\omega = e^{2\pi i/3}$. The lazy walk exhibits ballistic side peaks together with a pronounced central rest state peak, illustrating the three channel structure of the dynamics.\label{lazy_vs_standar_walk}
    }
\end{figure}
The lazy DTQW exhibits three distinct propagation branches, left, right, and a central peak associated with the rest subspace. Figure~\ref{lazy_vs_standar_walk} shows the probability distribution of the three state lazy walk. For comparison, a standard two state quantum walk is also shown, generated by a $2\times2$ Hadamard coin acting on $\{|L\rangle,|R\rangle\}$, following equation \eqref{wavefun_evolution_eqn} but with a two dimensional coin space, producing the familiar two ballistic peaks moving symmetrically away from the origin. In contrast, the three state lazy quantum walk generates a characteristic triplet structure with two ballistic side peaks together with a pronounced central peak associated with the rest state. The presence of this rest component alters both the symmetry and the overall dispersion of the walk, and forms the basic microscopic structure that the continuum limit must reproduce.

\subsection{Fourier Basis}
In order to analyse the continuum limit, it is convenient to work in the momentum representation, where the shift operator becomes diagonal and separable from the coin operator \citep{ambainis2001}. We therefore first transform both operators to Fourier space, before expressing them in the $\mathcal{SU}(3)$ Gell–Mann basis used for the continuum expansion. To obtain the momentum–space form of the walk, we apply the spatial Fourier transform:
\begin{equation}
\tilde{\boldsymbol{\psi}}(k,t)=\sum_x e^{ikx}\,\boldsymbol{\psi}(x,t),
\end{equation}
to the real space update rule implied by \eqref{wavefun_evolution_eqn}. 
Here $k$ is the dimensionless lattice wavenumber, related to the the physical momentum \(p\) via the de Broglie relation \(p=\hbar k\). Using the action of the shift operator in \eqref{shift-basic}, the evolution yields the local recurrence:
\begin{equation}
\boldsymbol{\psi}(x,t{+}1)
=
\mathbf{\hat{M}}_{+}\,\boldsymbol{\psi}(x{-}\Delta_x,t)
+ \mathbf{\hat{M}}_{0}\,\boldsymbol{\psi}(x,t)
+ \mathbf{\hat{M}}_{-}\,\boldsymbol{\psi}(x{+}\Delta_x,t),
\end{equation}
where $\mathbf{\hat{M}}_{-}= |L\rangle\langle L|\mathbf{\hat{C}}_G,  \quad \mathbf{\hat{M}}_{+}= |R\rangle\langle R|\mathbf{\hat{C}}_G$ and $M_0=| S\rangle\langle S|\,\mathbf{\hat{C}}_G$. Applying the Fourier transform gives:
\begin{equation}
\tilde{\boldsymbol{\psi}}(k,t{+}1)
=
\Bigl(e^{+ik\Delta_x}\mathbf{\hat{M}}_{+} \;+\; \mathbf{\hat{M}}_{0} \;+\; e^{-ik\Delta_x}\mathbf{\hat{M}}_{-}\Bigr)
\tilde{\boldsymbol{\psi}}(k,t)
\end{equation}
Defining the diagonal momentum-space shift operator
\begin{equation}
\tilde S(k)
= e^{-ik\Delta_x}|L\rangle\langle L| \;+\; |S\rangle\langle S| \;+\; e^{+ik\Delta_x}|R\rangle\langle R|,
\end{equation}
we therefore obtain the Fourier–space walk operator in the compact form
\begin{equation}
\tilde{\boldsymbol{\psi}}(k,t{+}1)
= \mathbf{\tilde S}(k)\,\mathbf{\hat{C}}_G\,\tilde{\boldsymbol{\psi}}(k,t).
\end{equation}
The conditional shift operator given by equation \eqref{shift-basic} becomes diagonal in the momentum basis:
\begin{equation}\label{shift-matrix}
\tilde S(k) =
\begin{pmatrix}
e^{-i k\Delta_x} & 0 & 0\\[4pt]
0 & 1 & 0\\[4pt]
0 & 0 & e^{i k\Delta_x},
\end{pmatrix}
\end{equation}
which shows explicitly that the \(|R\rangle\) component shifts forward, 
the \(|L\rangle\) component shifts backward, and the \(|S\rangle\) component remains localised.

The Grover coin, given by equation (\ref{grover-coin}) admits a simple spectral decomposition, which enables us to parameterise it as the exponential of a generator. We define $G=\mathbb{I}_3 - |u\rangle\langle u|$ with $|u\rangle$ being defined in equation (\ref{u_vector}), then
\begin{equation}\label{G_matrix}
G = 
\frac{1}{3}
\begin{pmatrix}
2 & -1 & -1 \\
-1 & 2 & -1 \\
-1 & -1 & 2
\end{pmatrix}
\end{equation}
A convenient parametrisation is given by
\begin{equation}\label{coin_as_exp}
\mathbf{\hat{C}}(\theta) = e^{-i\theta G}
= e^{-i\theta(\mathbb{I}_3 - |u\rangle\langle u|)}
= |u\rangle\langle u|
  + e^{-i\theta}\,(\mathbb{I}_3 - |u\rangle\langle u|).
\end{equation}
In this representation, the eigenstate corresponding to the uniform vector $|u\rangle$ has eigenvalue $+1$, while the two orthogonal Fourier modes acquire a common phase $e^{-i\theta}$. Thus, the coin operator $\hat{\mathbf{C}}(\theta)$ forms a smooth one parameter family that interpolates around the Grover operator. In particular, the standard Grover coin is recovered at the special value $\theta=\pi$, so that $\hat{\mathbf{C}}_G = \hat{\mathbf{C}}(\pi)$. This parametrisation introduces a tunable coin angle that controls the relative phase between the symmetric mode and the two transverse modes, and is therefore the natural small parameter for the continuum expansion. The parameter $\theta$ controls the strength of mixing between the three coin states.  In the continuum limit this coupling plays the role of an effective mass term. Small $\theta$ yields nearly ballistic propagation, while larger values increase local mixing and open a mass-like gap in the dispersion. The full walk operator in momentum space can be written as:
\begin{equation}
\mathbf{ \tilde U}(k, \theta) = \mathbf{\tilde S}(k)\,\mathbf{\mathbf{\hat{C}} }(\theta)
\end{equation}

\subsection{Gell Mann Representation}

We now express $\tilde S(k)$ and $\mathbf{\hat{C}}(\theta)$ in the 
$\mathcal{SU}$(3) Gell Mann basis.  This representation makes the underlying generators explicit and is the natural starting point for performing the continuum limit expansion of the walk operator in section \ref{Section_cont_limit}. 

For two state quantum walks, the two-dimensional coin operators can be decomposed in the Pauli basis, any $\mathcal{SU}$(2) operator can be written as a linear combination of Pauli matrices. In the lazy quantum walk, the coin space is three dimensional, so the natural generalisation is to use the eight Gell Mann matrices \(\{\lambda_i\}_{i=1}^8\), traceless Hermitian operators, which, together with the identity \(\mathbb{I}_3\), span the space of \(3\times3\) Hermitian operators \citep{gellmannmatrices}. The explicit form of each Gell Mann matrix can be found in Appendix \ref{app}. They satisfy the orthogonality relation \(\mathrm{Tr}(\lambda_i \lambda_j) = 2 \delta_{ij}\), so that any \(\mathcal{SU}(3)\) operator \(A \in \mathbb{C}^{3\times 3}\) can be expanded as
\begin{equation}
A = \frac{1}{3}\,\mathrm{Tr}(A)\, \mathbb{I}_3 + \frac{1}{2}\sum_{i=1}^8 
\mathrm{Tr}(A \lambda_i)\,\lambda_i .
\end{equation}
Having expressed both the shift and coin in a separable form in Fourier space, we
now rewrite these operators in the $\mathcal{SU}(3)$ Gell Mann basis. We first define the $\mathcal{SU}(3)$ generator:
\begin{equation}\label{J_z-matrix}
    J_z^{(3)} \;=\; \tfrac{1}{2}\lambda_3 + \tfrac{\sqrt{3}}.{2}\lambda_8
\;=\;
\mathrm{diag}(1,0,-1)
\end{equation}
Then the shift operator, given by equation \eqref{shift-matrix}, may be written as:
\begin{equation}\label{basic-shift}
\mathbf{\tilde S}(k)
= \exp\!\bigl[-\,i\,\Delta_x k\,(\mathbb{I}_x\otimes J_z^{(3)})\bigr].
\end{equation}
For notational simplicity, we omit tensor product identity operators $\mathbb{I}_x$ and $\mathbb{I}_c$ from this point onwards, writing $J_z^{(3)}$ and $G$ to denote the corresponding operators acting on the coin subspace.

The symmetric action of the Grover type coin is captured by the generator $G$ given by equation (\ref{G_matrix}). The generator $G$ may be written in the Gell Mann basis as:
\begin{equation}\label{G_matrix_GM}
G \;=\; \frac{2}{3}\mathbb{I}_3 - \frac{1}{3}\bigl(\lambda_1 + \lambda_4 + \lambda_6\bigr),
\end{equation}
which mixes the three internal states. 

The above representations are suitable to be used in the expansion of the unitary operator in the continuum limit (section \ref{Section_cont_limit}). Having established the basic formalism of the lazy DTQW, the next subsection introduces decoherence and outlines a model employed to incorporate it into the quantum walk framework.

\subsection{Decoherence in Quantum Walks}\label{intro_decoherence}

Decoherence is usually described as a non-unitary evolution using the density matrix formalism to describe the mixed states that arise.  Simple decoherence models for quantum walks were introduced in \cite{Kendon-useful}. Here, we use a similar simple Lindblad master equation to examine the basic features of decoherence in the lazy quantum walk, providing a baseline against which more complex decoherence mechanisms can be compared. This approach assumes a Markovian open system description, in which the environment has a short correlation time and no memory effects are retained. The Lindblad master equations provides dynamics that are completely positive and trace preserving, ensuring that the density operator remains physical at all times, regardless of the step size, as we take the limits \citep{deco_in_1d}. The coupling to the environment is treated phenomenologically through a single decoherence strength~$\gamma$, without specifying a microscopic bath model or spectral density. Initial system-environment correlations are neglected, and the system follows a weak decoherence regime \citep{BreuerPetruccione2002}. 

All the operators act in the same Hilbert space \( \mathcal{H}_{xc} = \mathcal{H}_x \otimes \mathcal{H}_c \) as the unitary quantum walk. The system is described by a time dependent density operator, expressed in the computational basis as
\begin{equation}
\boldsymbol{ \hat\rho}(t) = \sum_{x,c}\sum_{x',c'} \rho_{x,c}^{x',c'}(t)|x,c\rangle\langle x',c'|.
\end{equation}
The density operator is Hermitian, positive semi-definite, of unit trace, and bounded. It governs the evolution of both the internal coin and spatial degrees of freedom. 
A single timestep discrete time evolution, $\Delta_t$, is governed by the map:
\begin{equation}\label{most-basic-discrete}
\boldsymbol{\hat \rho}(t+\Delta_t) = (1-\gamma)\hat{\mathbf{U}}\boldsymbol{\hat \rho}(t)\hat{\mathbf{U}}^\dagger + \gamma\sum_j \mathbb{P}_j\hat{\mathbf{U}}\boldsymbol{\hat \rho}(t)\hat{\mathbf{U}}^\dagger\mathbb{P}_j^\dagger,
\end{equation}
where \( \gamma \in [0,1] \) is the probability of decoherence per full step, and the operator \( \tilde{\mathbf{U}} \) is the full unitary evolution operator. This decoherence model was reviewed in \cite{KENDON_2007} and evolved numerically for various choices of $\mathbb{P}_j$. This is the infinite temperature Lindblad form, whose stationary state is the maximally mixed state and treats all energy levels symmetrically. 

Rewriting in standard form, the Lindblad CPTP map is given by
\begin{equation}\label{discrete lindblad-equation}
    \boldsymbol{\hat\rho} (t + \Delta_t) =  \mathbf{\hat{U}}\boldsymbol{\hat{\rho}}(t)\mathbf{\hat{U}} + \gamma \sum_j \big(L_{j}  \boldsymbol{\hat\rho}(t) {L}_{j}^{\dagger} - \frac{1}{2}\big\{ L_{j}^{\dagger}L_{j}, \boldsymbol{\hat\rho}(t)\big\})
 \end{equation}
The decoherence is modelled as projective noise using jump operators constructed as
\begin{equation}\label{jump-op}
L_j = \mathbb{P}_j \, \mathbf{\hat U},
\end{equation}
where the projectors \(\{\mathbb{P}_j\}\) specify the particular decoherence channel. 
For coin decoherence, the projectors act on the coin space, so that off-diagonal coherences between coin states are suppressed while diagonal populations are preserved.  For spatial decoherence, the projectors act on the position space, which destroys coherence between different spatial locations while leaving the coin degree of freedom unaffected. In both cases the decoherence parameter \(\gamma\) controls the strength of the noise, 
setting the probability of a projective decoherence event at each timestep. The 
projectors satisfy the usual relations
\begin{equation}
\mathbb{P}_j^\dagger = \mathbb{P}_j, \quad 
\mathbb{P}_j^2 = \mathbb{P}_j, \quad 
\sum_j \mathbb{P}_j = \mathbb{I}.
\end{equation}
Decoherence modifies the characteristic spreading peaks of the unitary quantum walk, and potentially induces a crossover to classical behaviour, depending on the model, strength and type of environmental coupling \citep{KENDON_2007}.  The continuum limit of equation \eqref{discrete lindblad-equation} may be taken after specifying the structure of the unitary operator and decoherence model. We first recap the effects of both coin and spatial decoherence on the discrete-time lazy quantum walk.
    \begin{figure}[!ht]
\centering
    \subfloat[Probability distribution for the case of coin decoherence over several values of $\gamma$ for coin angle $\theta = \frac{\pi}{2}$. ]{\includegraphics[width=0.45\linewidth]{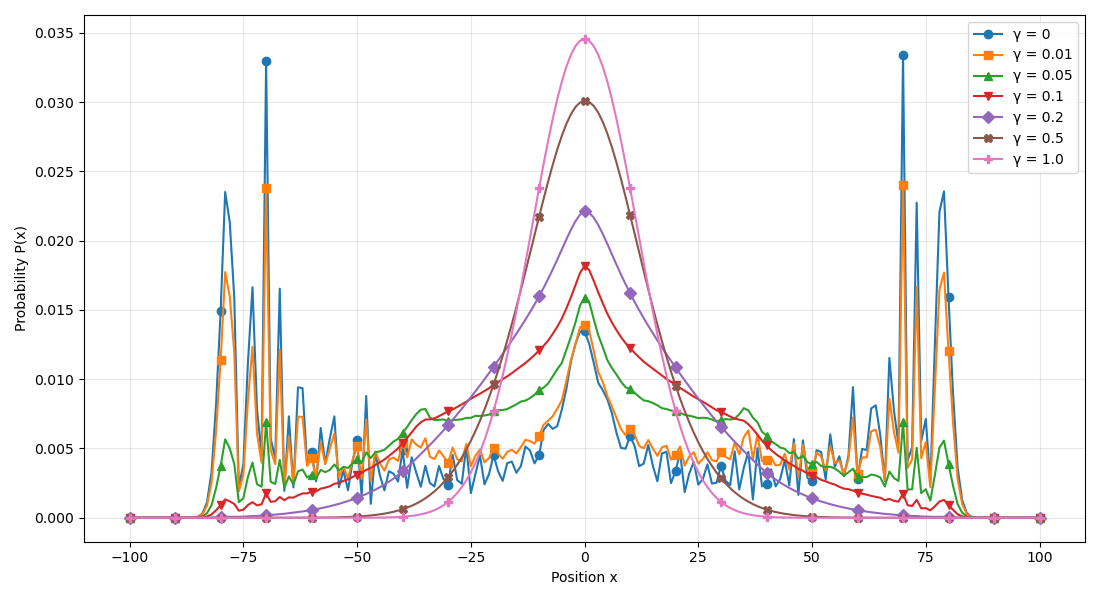}}
\hfil
    \subfloat[Probability distribution for the case of spatial decoherence over several values of $\gamma$ for coin angle $\theta = \frac{\pi}{2}$.]{\includegraphics[width=0.45\linewidth]{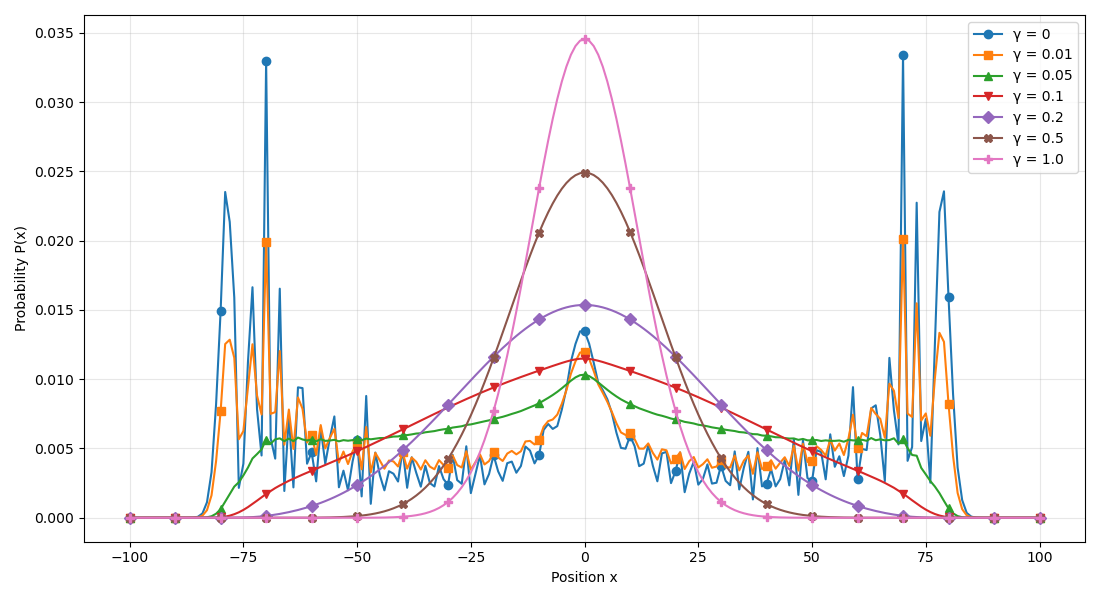}}

    \subfloat[Probability distribution for the case of coin decoherence over several values of $\gamma$ for coin angle $\theta = \pi$.]{\includegraphics[width=0.45\linewidth]{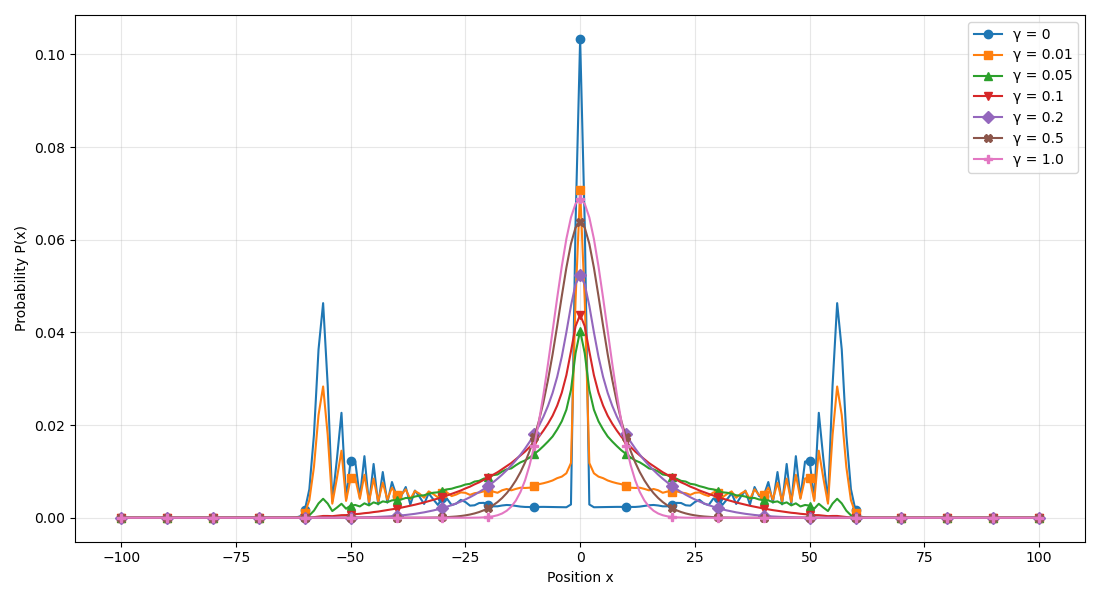}}
\hfil
    \subfloat[Probability distribution for the case of spatial decoherence over several values of $\gamma$ for coin angle $\theta = \pi$.]{\includegraphics[width=0.45\linewidth]{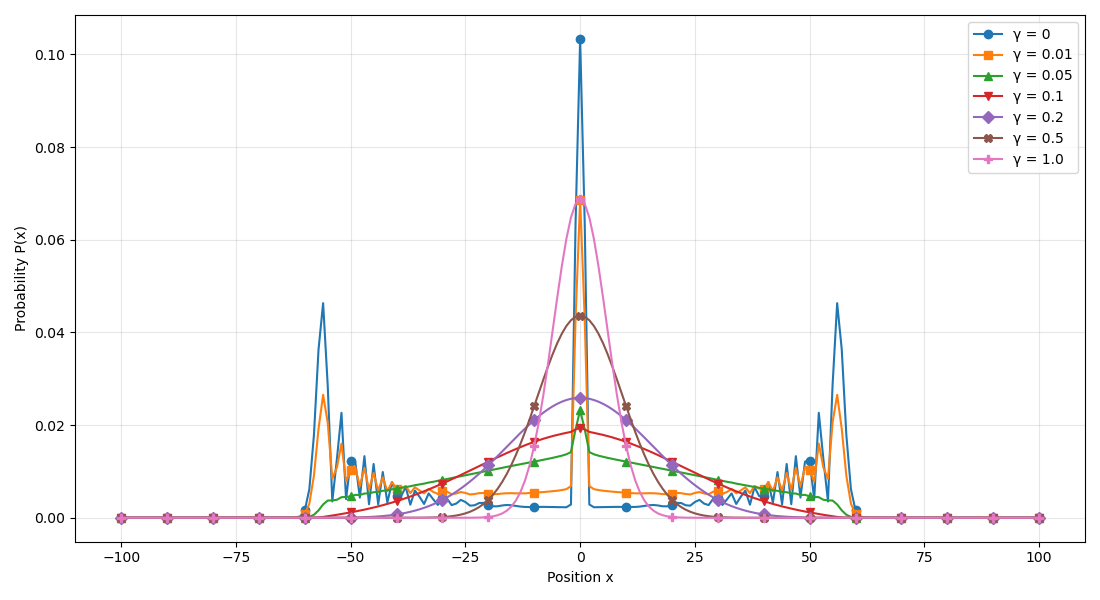}}

\caption{Effect of dephasing on the lazy DTQW for different coin angles $\theta$ and decoherence strengths $\gamma$. Panels (a,c) show coin dephasing and panels (b,d) show spatial dephasing, for $\theta=\pi/2$ (top row) and $\theta=\pi$ (bottom row). The initial state is a localised position state with a Fourier symmetric coin superposition, $|\psi_0\rangle = |0\rangle \otimes \frac{1}{\sqrt{3}}\bigl(|L\rangle + \omega\,|S\rangle + \omega^{2}\,|R\rangle\bigr)$, where $\omega = e^{2\pi i/3}$.
}
    \label{fig:deco_compare}
    \end{figure}

Figure~\ref{fig:deco_compare} compares the effects of coin and spatial dephasing on the lazy DTQW for several values of the decoherence strength~$\gamma$ and for different coin angles~$\theta$. Coin dephasing, shown in panels~(a,c), selectively damps the off-diagonal coherences between the internal coin states $\{|L\rangle,|S\rangle,|R\rangle\}$ while leaving spatial coherences intact. In contrast, spatial dephasing, shown in panels~(b,d), suppresses off-diagonal density matrix elements $\rho(x,x')$ for $x\neq x'$. As~$\gamma$ increases, the ballistic side peaks collapse more rapidly than in the coin dephased case and the probability distribution evolves more rapidly toward a broad, approximately Gaussian profile centred at the origin. This transition occurs already at relatively small values of~$\gamma$, reflecting the sensitivity of ballistic transport to spatial coherence. For the spatially dephased case, the probability distribution seems to flatten out at low~$\gamma$ values before tending towards the Gaussian profile. This is consistent with the behaviour of of the two state quantum walk under projective decoherence. In the limit $\gamma \to 1$, both decoherence mechanisms yield a similar Gaussian distribution, corresponding to a fully classical diffusive regime. Comparing different coin angles further reveals the role of the coin in controlling transport and localisation. For $\theta = \pi$, the unitary walk exhibits a pronounced central peak, indicating that the coin strongly favours population of the rest state and suppresses propagation. For $\theta = \pi/2$, the coin induces stronger mixing between the propagating and non-propagating components, resulting in a less pronounced central peak and a higher probability of reaching the ballistic fronts. This behaviour persists under both decoherence mechanisms, although spatial dephasing again accelerates the transition toward a Gaussian profile. Overall, increasing $\theta$ modifies the internal coin mixing and interference structure of the walk, leading to systematically reduced spatial spreading and enhanced localisation near the origin.

\section{Continuum Limit of the Lazy Quantum Walk with Decoherence}
\label{Section_cont_limit}

The previous section introduced the discrete Lindblad formulation of the lazy quantum walk with decoherence. In this section, that framework is applied to specific choices of decoherence models, and their continuum limits are derived to identify the effective dynamics simulated by each case. The following subsections derive the continuous spacetime limit of lazy quantum walks with both coin and spatial decoherence. Although both models share the same underlying unitary dynamics, they differ in the structure of the Lindblad term, leading to a different effective PDE in the limit. 

\subsection{Scaling and Parameterisation}

The continuum limit of equation \eqref{most-basic-discrete} is obtained by Taylor expanding the discrete Lindblad evolution operator in powers of a small parameter~$\varepsilon$. This parameter simultaneously scales the lattice spacing, timestep, and coin angle. The continuum limit is then triggered by taking~$\varepsilon \rightarrow 0$. Following the standard approach \citep{dimolfetta2021}, we set:
\begin{equation}\label{scaling-params}
\begin{split}
& \Delta_t = \varepsilon\\
& \Delta_x = \varepsilon\\
& \theta   = \varepsilon\,\bar\theta
\end{split}
\end{equation}
where $\bar\theta$ is a constant and remains finite in the limit. The decoherence rate is also weakly scaled:
\begin{equation}\label{gamma-scalih}
    \gamma = \varepsilon\,\bar\gamma,
\end{equation}
This scaling corresponds to the weak decoherence regime, in which the noise acts only as a small perturbation on the otherwise coherent dynamics. The constant $\bar{\gamma}$ remains finite in the limit and represents the effective decoherence rate per unit time step. If instead, the unscaled rate $\gamma$ were kept fixed as $\varepsilon \to 0$, the decoherence would remain strong in the continuum limit, causing the Lindbladian term to dominate over the unitary advection and mixing contributions. Such a regime leads to rapid relaxation and trivial stationary behaviour, and is therefore not considered further here. With the scaling defined above, we can now expand the unitary evolution operator to obtain the effective Hamiltonian governing the coherent part of the dynamics.

\subsection{Unitary Expansion and Effective Hamiltonian}
\label{sec:unitary-limit}

Before introducing decoherence, the unitary part of the continuum limit is derived. This component is common to both models considered below, since each decoherence mechanism employs the same underlying shift and coin operators. The differences between models arise solely through the choice of projectors acting within the Lindblad term, while the unitary evolution remains identical. The discrete unitary step operator is defined as:
\begin{equation}
\mathbf{\tilde U}(k) = \mathbf{\tilde S}(\varepsilon, k)\,\mathbf{\mathbf{\hat{C}}}(\varepsilon,\bar{\theta}),
\end{equation}
where both the shift and coin operators are expressed as exponentials of $\mathcal{SU}(3)$ generators. The coin operator given by equation (\ref{coin_as_exp}) using the generator given by equation (\ref{G_matrix}) may now be expanded around $\varepsilon$, as:
\begin{equation}
    \mathbf{ \mathbf{\hat{C}}}(\varepsilon, \bar{\theta}) = e^{-i\varepsilon\bar{\theta} G} \approx \mathbb{I}_3 - i\varepsilon\bar{\theta} G - \frac{\varepsilon^2 \bar{\theta}^2}{2}G^2 + \mathcal{O}(\varepsilon^3)
\end{equation}
The shift operator, given by equation (\ref{basic-shift}), can then be expanded around $\varepsilon$ as:
\begin{equation}
    \mathbf{\tilde{S}}(\varepsilon, k) = e^{-i\varepsilon k J_z^{(3)}} \approx \mathbb{I}_3 - i\varepsilon k J_z^{(3)} - \frac{\varepsilon^2 k^2}{2} \big(J_z^{(3)}\big)^{2} + \mathcal{O}(\varepsilon^3)
\end{equation}
 Evaluating the first order expansion of 
$\mathbf{\tilde U}(k)=\mathbf{\tilde S}(\varepsilon,k)\mathbf{\mathbf{\hat{C}}}(\varepsilon,\bar{\theta})$ small values of $k$ and $\bar{\theta}$, the unitary operator can be expanded around the identity to find the Hamiltonian $\hat{\mathbf{H}}_{phys}$:
\begin{equation}
\mathbf{\tilde U}(k)
\simeq
\mathbb{I}_3
- i\varepsilon\,\hat{\mathbf{H}}_{\text{phys}}
+ \mathcal{O}(\varepsilon^2),
\end{equation}
where $\hat{\mathbf{H}}_{\text{phys}}$ is the effective Hamiltonian governing the continuum time evolution. The unitary Hamiltonian is then found to be:
\begin{equation}
\hat{\mathbf{H}}_{\text{phys}}(k) 
= k\,J_z^{(3)} + \bar{\theta}\,G
= k\!\left(\tfrac{1}{2}\lambda_3+\tfrac{\sqrt{3}}{2}\lambda_8\right)
+\frac{\bar{\theta}}{3}\big(\lambda_1+\lambda_4+\lambda_6\big),
\end{equation}
where $J_z^{(3)}$ generates directional propagation along the lattice and $G$ produces symmetric mixing between the three internal states. To obtain a real space description, we invert the Fourier representation of the walk
and take the continuum limit, in which the quasi-momentum $k$
enters only through smooth phase factors generated by lattice translations. Expanding these phase factors to leading order in $k$ and transforming back to
position space maps the momentum dependence of the generator onto spatial derivatives acting on the wavefunction. As a consequence, the momentum variable is represented in real space by the
differential operator $k \rightarrow -\,i\,\partial_x$, yielding the effective real space Hamiltonian
\begin{equation}\label{H_phys}
\hat{\mathbf{H}}_{\text{phys}} = -i\,J_z^{(3)}\,\partial_x + \bar{\theta}\,G,
\end{equation}
so that the corresponding continuum dynamics satisfy
\begin{equation}\label{conitnuum-unitary-limit}
\,\partial_t\psi(x,t)
= -i \hat{\mathbf{H}}_{\text{phys}}\psi(x,t),
\end{equation}
which describes ballistic advection along the lattice direction accompanied by local symmetric mixing in the internal state space. In the massless limit $\bar{\theta}=0$, the Hamiltonian generates purely linear propagation of the left and right moving components with dispersion $\omega=\pm k$, while the lazy channel $\ket{S}$ remains dynamically decoupled. The resulting dynamics are formally analogous to the one dimensional propagation of a free, massless spin-1 field, with the two propagating components playing the role of helicity eigenmodes. In this sense, the unitary $\mathcal{SU}(3)$ walk provides an analogue of photon-like propagation in one spatial dimension \citep{bialynickibirula2005photonwavefunction}. It is important to note, however, that this analogy is kinematic rather than exact. Unlike the electromagnetic field, the present model does not possess gauge invariance or an associated transversality constraint, and the additional internal component is therefore retained as a physical degree of freedom rather than being removed by a gauge redundancy. The lazy channel $\ket{S}$ thus acts as a longitudinal or rest mode that extends the minimal two component Dirac structure to a three component multiplet, rather than representing an unphysical gauge degree of freedom.

In terms of field theories, the parameter $\theta$ plays the role of an effective mass. This interpretation follows directly from the structure of the Hamiltonian in~\eqref{H_phys}. The kinetic term $-iJ_z^{(3)}\partial_x$ generates advection of the left and right moving components, while the $\bar{\theta},G$ term induces local mixing in coin space that couples these propagation channels. In two-component quantum walks, such mixing terms are responsible for opening an effective mass gap in the dispersion relation. The same mechanism appears here in its $\mathcal{SU}(3)$ generalisation, as the operator $G$ couples the three internal states symmetrically and acts as the analogue of a mass generating interaction. Consequently, the parameter $\bar{\theta}$ controls the strength of internal-state coupling and determines the opening of an effective mass gap. Small values of $\bar{\theta}$ yield nearly massless, strongly dispersive behaviour with sharply separated ballistic branches, while larger $\theta$ enhances mixing and progressively suppresses coherent propagation, leading to a flattening of the dispersion relation. This trend is visible from the $\gamma = 0$ lines in Fig.~\ref{fig:deco_compare}, where increasing the effective mass parameter $\bar{\theta}$ suppresses ballistic propagation and reduces the maximum distance reached at fixed $N$, whereas smaller $\bar{\theta}$ corresponds to lighter, more freely propagating dynamics. The Hamiltonian $\hat{\mathbf{H}}_{\text{phys}}$ therefore represents the minimal $\mathcal{SU}(3)$ embedding of a Dirac-type continuum structure, with $J_z^{(3)}$ supplying the kinetic generator and $G$ providing the mass coupling.

This structure defines the universal unitary backbone of the model: all differences between the decoherence channels developed below arise exclusively from how their respective Lindblad projectors act on this shared $\mathcal{SU}(3)$ Hamiltonian, while the underlying effective field theory description of the unitary dynamics remains the same.

\subsection{Including Decoherence}

We now incorporate decoherence into the discrete time evolution. The dynamics are governed by the Lindblad CPTP map given in equation~\eqref{discrete lindblad-equation}, with jump operators defined in equation~\eqref{jump-op}. Applying the chosen scaling with respect to the small parameter~$\varepsilon$, we obtain the discrete Lindblad update rule, which we will subsequently expand to derive its continuum limit:
\begin{equation}
\boldsymbol{\hat\rho}(t+\varepsilon)
= \boldsymbol{\hat\rho} (t ) -i\varepsilon[\mathbf{\hat{H}},\rho(t)]
+ \varepsilon\bar\gamma\sum_j
\!\left(
\mathbb{P}_j\mathbf{\tilde U}\boldsymbol{\hat\rho}(t)\mathbf{\tilde U}^\dagger \mathbb{P}_j^\dagger
-\tfrac{1}{2}\big\{\mathbf{\tilde U}^\dagger \mathbb{P}_j^\dagger \mathbb{P}_j\mathbf{\tilde U},\boldsymbol{\hat\rho}(t)\big\}
\right)
\end{equation}
where the projectors $\mathbb{P}_j$ depend on whether decoherence acts on the coin or position subspace.  The same unitary generator $\hat{\mathbf{H}}_{\text{phys}}$ derived above appears in both cases. The following subsections derive the continuum limit for each model, emphasising that only the Lindblad term differs.

\subsubsection{Coin Decoherence}
\label{sec:coin-decoh}

We begin by considering the case in which decoherence acts solely on the coin subspace. In this setting, the projectors that define the decoherence channel act within the internal coin basis:
\begin{equation}\label{projectors}
\mathbb{P}_{L}=\ket{L}\bra{L},\qquad
\mathbb{P}_{S}=\ket{S}\bra{S},\qquad
\mathbb{P}_{R}=\ket{R}\bra{R}.
\end{equation}
These projectors are substituted into the discrete Lindblad equation and we expand in~$\varepsilon$ to first order. We take the continuum limit and  and retain the leading terms in~$\varepsilon$. The real space description is obtained by inverting the Fourier representation, and consequently the momentum is replaced by the differential operator $k\mapsto -\,i\,\partial_x$ in $\hat{\mathbf{H}}_{\text{phys}}$. For clarity, it is convenient to introduce the projector onto the diagonal in the coin basis
\begin{equation}
\Pi_c[\boldsymbol{\rho}] \;=\; \mathrm{diag}(\boldsymbol{\rho}),
\end{equation}
which keeps only the diagonal coin populations in the $\{\ket{L},\ket{S},\ket{R}\}$ basis. The dephasing channel can then be written as
\begin{equation}
D_c[\boldsymbol{\rho}]
\;=\;
\boldsymbol{\rho} - \Pi_c[\boldsymbol{\rho}],
\label{eq:Dc-def}
\end{equation}
so that $D_c[\boldsymbol{\rho}]$ vanishes on populations and acts solely on coherences between the internal states. With this notation, the full coarse grained continuum equation reads:
\begin{equation}
\partial_t \boldsymbol{\rho}(x,t)
\;=\;
\,[\boldsymbol{\rho}(x,t), \partial_xJ_z^{(3)}]
\;-\; i\,\bar{\theta}\,[G,\,\boldsymbol{\rho}(x,t)]
\;-\;\bar{\gamma}\,D_c[\boldsymbol{\rho}(x,t)],
\label{eq:full-continuum-PDE-coin-deco}
\end{equation}
where the coherent part is expressed in the Gell Mann basis via the fixed $\mathcal{SU}(3)$ generators $J_z^{(3)}$ and $G$, given by equations~\eqref{J_z-matrix} and~\eqref{G_matrix_GM} respectively. The first commutator in equation~\eqref{eq:full-continuum-PDE-coin-deco} generates directional streaming of the three coin channels (left, rest, right), while the second implements their symmetric local mixing. The final term, $-\bar{\gamma}\,D_c[\boldsymbol{\rho}]$, damps only the off diagonal coin coherences, leaving the coin populations unchanged at leading order.

Equation~\eqref{eq:full-continuum-PDE-coin-deco} is linear, first order in time, and local in~$x$, with the spatial derivative entering only through the commutator with $J_z^{(3)}$. It preserves Hermiticity and trace, and for $\bar{\gamma}\!\ge\!0$ generates a CPTP semigroup on the coin space at each spatial point. The unitary part retains the full $\mathcal{SU}(3)$ structure of the lazy walk, while the dephasing produces a uniform decay of the six off diagonal Gell–Mann components $\{\lambda_{1},\lambda_{2},\lambda_{4},\lambda_{5},\lambda_{6},\lambda_{7}\}$ and leaves the diagonal sector $(\lambda_3,\lambda_8)$ intact. Consequently, the continuum dynamics preserve the underlying streaming structure while progressively reducing the interference between the left, rest, and right channels in space and time. The absence of diagonal contributions in $D_c$ indicates that the decoherence channel does not directly modify the populations of the coin states at leading order, but instead damps their mutual phase coherences. While the unitary evolution may induce oscillations of the coin populations depending on the initial state, the action of the decoherence term selectively suppresses the off-diagonal components of the coin density matrix. Physically, this implies that probability weight is redistributed between internal levels solely through coherent dynamics, whereas interference between coin states is progressively lost. As a result, the coherent oscillations characteristic of the unitary walk are gradually attenuated, and the system evolves towards an effectively classical mixture in the coin subspace, providing a microscopic mechanism for the emergence of classical behaviour in the macroscopic limit.

\begin{figure}
    \centering
\includegraphics[width=0.8\linewidth]{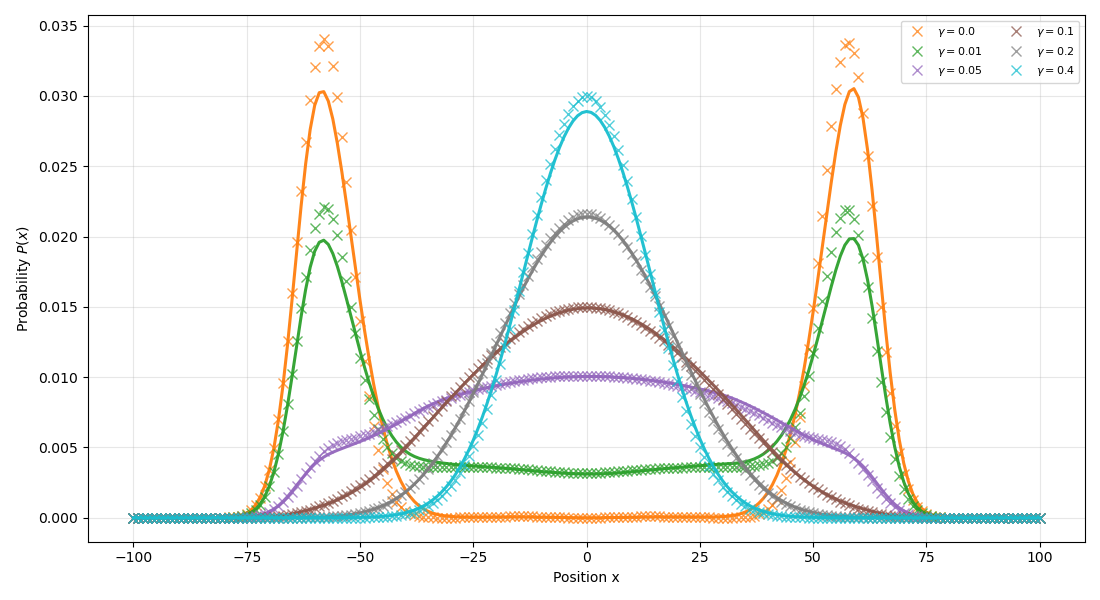}
    \caption{Discrete time quantum walk versus derived continuum master equation for coin dephasing. Final position probability $P(x,t)$ after $N=100$ steps of the lazy three state quantum walk with $\theta=\pi/2$, shown for several decoherence rates $\gamma$. Solid lines show the solution of the corresponding derived continuum Lindblad equation. The initial state is a spatially delocalised Gaussian wavepacket of width $\sigma_0 = 5$ with a Fourier-symmetric coin superposition $(1,\omega,\omega^2)/\sqrt{3}$, where $\omega = e^{2\pi i/3}$.
}

    \label{cont_vs_discrete_coin_deco}
\end{figure}

Figure~\ref{cont_vs_discrete_coin_deco} shows the position probability distributions obtained from DTQW and from the corresponding derived continuum equation for several values of the coin dephasing parameter $\gamma$. The figure demonstrates very good agreement between the DTQW and the continuum Lindblad dynamics throughout the range of weak to moderate dephasing considered here ($0 \leq \gamma \leq 0.4$), which is the weak decoherence regime in which the continuum master equation is derived. Within this range, the continuum solution accurately reproduces both the overall shape and the width of the position distribution produced by the discrete walk. In particular, the crossover from ballistic, multi peaked propagation at weak dephasing to a smoother, more centrally concentrated profile as the dephasing strength increases is captured faithfully. This agreement provides strong evidence that the DTQW with coin dephasing constitutes a consistent and controlled discretisation of the target continuum open quantum dynamics in its domain of validity.

\subsection{Spatial Decoherence}

We now consider the complementary case in which decoherence acts on the spatial subspace rather than on the coin. The discrete evolution is still governed by the Lindblad map in equation~\eqref{discrete lindblad-equation} with jump operators of the form~\eqref{jump-op}, but the projectors now resolve the position basis:
\begin{equation}
\mathbb{P}_x \;=\; \ket{x}\bra{x}\,\otimes\,\mathbb{I}_c,
\label{eq:Px-def}
\end{equation}
so that each Lindblad operator isolates a single lattice site while leaving the internal (coin) degree of freedom untouched,
\begin{equation}
L_x \;=\; \mathbb{P}_x\,\tilde{\mathbf{U}}.
\label{eq:Lx-def}
\end{equation}
We assume the same weak coupling continuum scaling as in the coin-decoherence case, given by equations~\eqref{scaling-params} and~\eqref{gamma-scalih}. We then expand the discrete Lindblad step to first order in~$\varepsilon$. The discrete Lindblad map is again expanded to first order in~$\varepsilon$ and the continuous spacetime limit is taken. Again, the real space description is obtained by inverting the Fourier representation, and consequently the momentum is replaced by the differential operator $k\mapsto -\,i\,\partial_x$ in $\hat{\mathbf{H}}_{\text{phys}}$. To make the structure of the dissipator explicit, it is convenient to view $\boldsymbol{\rho}$ in the position representation, with matrix elements
\begin{equation}
\boldsymbol{\rho}_{cc'}(x,x',t)
\;=\;
\bra{x,c}\,\boldsymbol{\rho}(t)\,\ket{x',c'}
\end{equation}
We introduce the projector onto the diagonal in position space,
\begin{equation}
\big(\Pi_x[\boldsymbol{\rho}]\big)_{cc'}(x,x',t)
\;=\;
\delta_{x,x'}\,\boldsymbol{\rho}_{cc'}(x,x',t),
\label{eq:Pix-def}
\end{equation}
which removes all spatial off-diagonal components while preserving the full internal (coin) structure at each site. In analogy with the coin dephasing map~\eqref{eq:Dc-def}, we define
\begin{equation}
D_x[\boldsymbol{\rho}]
\;=\;
\boldsymbol{\rho} - \Pi_x[\boldsymbol{\rho}],
\label{eq:Dx-def}
\end{equation}
so that $D_x[\boldsymbol{\rho}]$ vanishes on the spatial populations (the blocks with $x=x'$) and acts only on coherences between distinct lattice sites. The dissipator in equation~\eqref{eq:spatial-deph-PDE} then has matrix elements
\[
\big(-\bar\gamma\,D_x[\boldsymbol{\rho}]\big)_{cc'}(x,x',t)
=
\begin{cases}
0, & x=x',\\[2pt]
-\bar\gamma\,\boldsymbol{\rho}_{cc'}(x,x',t), & x\neq x'.
\end{cases}
\]
With this notation, the full coarse grained continuum equation reads:
\begin{equation}\label{eq:spatial-deph-PDE}
\partial_t \boldsymbol{\rho}(x,t)
\;=\;
\,[\boldsymbol{\rho}(x,t), \partial_xJ_z^{(3)}]
\;-\; i\,\bar{\theta}\,[G,\,\boldsymbol{\rho}(x,t)]
\;-\;\bar{\gamma}\,D_x[\boldsymbol{\rho}(x,t)],
\end{equation}
where $\hat{\mathbf{H}}_{\text{phys}}$ is given in equation~\eqref{H_phys} and the spatial dephasing channel $D_x$. Here, the first two terms correspond to the density matrix form of equation (\ref{conitnuum-unitary-limit}), and the final term gives the spatial dissipator. Equation~\eqref{eq:spatial-deph-PDE} is again linear, first order in time, and local in $t$, with the coherent part identical to the coin-decoherence case and the dissipative part now acting purely in the spatial sector. It preserves Hermiticity and trace and, for $\bar\gamma\!\ge\!0$, generates a CPTP semigroup on the full coin–position Hilbert space. The Hamiltonian term $-i[\hat{\mathbf{H}}_{\text{phys}},\boldsymbol{\rho}]$ produces directional streaming and local coin mixing and is the density matrix form of equation (\ref{conitnuum-unitary-limit}). The dephasing channel $-\bar\gamma\,D_x[\boldsymbol{\rho}]$ leaves all on site density matrices $\boldsymbol{\rho}(x,x,t)$ invariant and drives an exponential decay of the off-diagonal blocks $\boldsymbol{\rho}(x,x',t)$ with $x\neq x'$.

Physically, this implies that the decoherence channel acts locally in position space and does not directly transfer probability between lattice sites or modify the internal coin populations at a given site, but instead suppresses coherent superpositions between different positions. As a consequence, although the walk retains its local streaming dynamics, the loss of long-range spatial coherence alters the interference mechanisms responsible for ballistic transport. The resulting evolution therefore differs qualitatively from the fully unitary walk, leading to a redistribution of probability in space. In the long-time limit, the state approaches a classical mixture in the position basis, providing a microscopic mechanism for the emergence of spatially classical transport while still resolving the internal coin degrees of freedom at each site. This makes clear why spatial decoherence is typically undesirable when quantum properties are required: dephasing in the position basis destroys the long-range interference that enables quantum advantage, leaving only the coin subspace quantum but removing the mechanism that produces ballistic transport.

\begin{figure}
    \centering
    \includegraphics[width=0.8\linewidth]{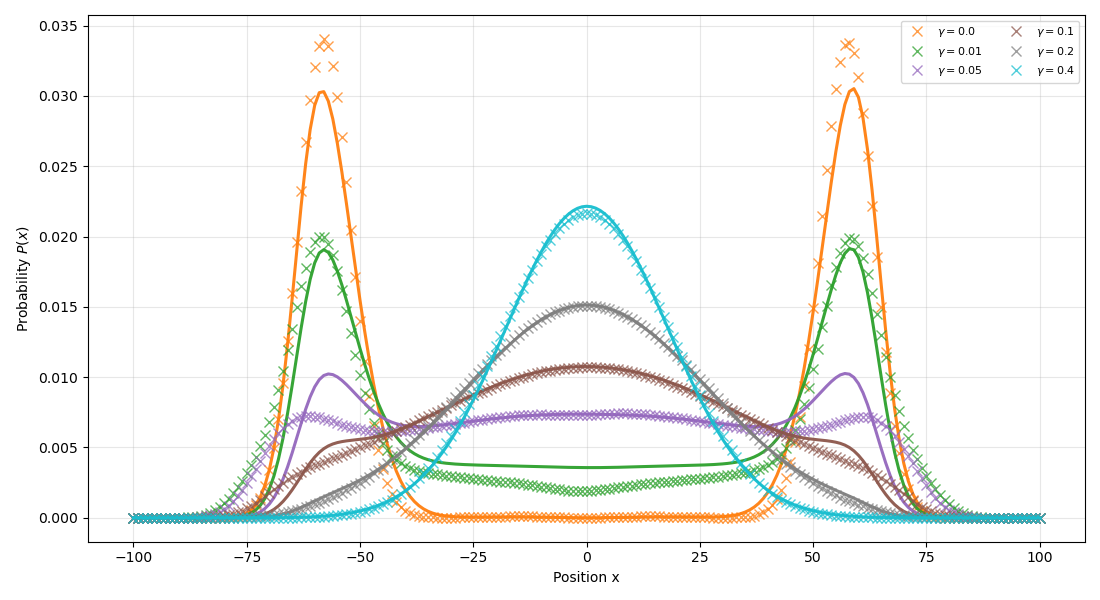}
    \caption{DTQWs derived continuum master equation for spatial dephasing. Final position probability $P(x,t)$ after $N=100$ steps of the lazy three state quantum walk with $\theta=\pi/2$, shown for several decoherence rates $\gamma$. Solid lines show the solution of the corresponding derived continuum Lindblad equation with spatial dephasing. The initial state is a spatially delocalised Gaussian wavepacket of width $\sigma_0 = 5$ with a Fourier-symmetric coin superposition $(1,\omega,\omega^2)/\sqrt{3}$, where $\omega = e^{2\pi i/3}$.}
    \label{cont_vs_discrete_spatial_deco}
\end{figure}

Figure \ref{cont_vs_discrete_spatial_deco} shows the final position probability distribution obtained from the DTQW with spatial dephasing (markers) compared to the corresponding continuum Lindblad equation (solid lines), for several decoherence strengths $0 \leq \gamma \leq 0.4$. Again, we restrict the range of gamma between these values as the continuum equation is derived under the weak decoherence assumption. As the dephasing rate increases, the dynamics smoothly interpolate from a ballistic, interference dominated regime with pronounced side peaks to a broader, more diffusive profile concentrated around the origin. The continuum model accurately reproduces this overall transition, including the suppression of the ballistic fronts and the emergence of a central peak at larger $\gamma$ values. Small quantitative differences are visible at low decoherence, particularly in the outer regions of the distribution, where lattice scale interference effects remain more pronounced in the discrete walk than in the continuum description. This slightly reduced agreement compared to the coin-dephasing case is expected, since spatial dephasing acts directly on nonlocal coherences between lattice sites, which are more sensitive to coarse graining in the continuum limit. Overall, the plot confirms that the continuum equation captures the dominant transport behaviour induced by spatial decoherence, while smoothing out fine interference features of the underlying discrete dynamics.

\section{Discussion and Conclusion}\label{discussion-sec}

We have derived the continuum limit of the one dimensional lazy open quantum walk for both coin and spatial decoherence models. The unitary part of the evolution generates advection and local mixing governed by the coin, while the dissipative part captures the distinct physical effects of each noise channel. In the continuum limit, the unitary structure remains identical and only the dephasing operator changes form, revealing how different microscopic decoherence mechanisms manifest at the macroscopic level. 

The coin dephasing Lindblad term removes off diagonal elements in the coin subspace while leaving spatial coherences between lattice sites intact. Physically, this results in pure coin dephasing. Local internal superpositions decay, but population transport remains coherent until diffusion dominates at high $\gamma$ values. This behaviour is in good qualitative agreement with early numerical studies of decoherence in the two state DTQW, which showed that coin noise suppresses interference without immediately destroying ballistic transport \citep{Paz_Lopez_deco, Brun_Ambainis, Love2005FromDT, Alberti_2014}. By contrast, spatial decoherence mechanisms were found to rapidly suppress long-range coherence and induce diffusive behaviour \citep{ROMANELLI2005137}, a qualitative distinction that is also observed here. Crucially, the numerical plots demonstrate good agreement between the DTQW and the derived continuum dynamics across the explored parameter regimes. This agreement confirms that the DTQW provides a faithful discretisation of the target continuum quantum field theory. The continuum equation represents the macroscopic model of interest, while the underlying discrete dynamics form a controlled numerical scheme for its simulation. Establishing this correspondence justifies the use of discrete quantum walks as simulators of continuum quantum transport equations and motivates the systematic derivation of the continuum limit \citep{Molfetta_Debbasch_limits_and_sym}. Spatial dephasing, by contrast, suppresses coherences between different lattice sites while leaving on-site populations unaffected. As a result, phase correlations decay exponentially in space, progressively eliminating long-range interference effects and driving the dynamics toward an incoherent, effectively classical transport regime. This behaviour aligns with previous numerical observations of spatially dephased quantum walks \cite{Tude_2022b}. Together, these results show that coin and spatial dephasing impose qualitatively different transport characteristics. Coin dephasing preserves coherent advection with local phase damping, while spatial dephasing removes all spatial coherence and yields incoherent propagation that quickly reaches a classical, diffusive regime. The continuum framework developed here thus bridges microscopic Lindblad dynamics with the macroscopic behaviour observed in open quantum walk simulations.

The analytical structure of these limits may also be useful for future hydrodynamic applications. The recent work of \cite{Au_Yeung_2025} illustrates how quantum walks can emulate nonlinear fluid dynamics by reconstructing hydrodynamic quantities from the collective behaviour of quantum walkers in a smoothed particle framework. This provides a clear example of how quantum walk dynamics can generate effective nonlinearities at the macroscopic level. While the present model captures only advective motion with local phase damping, it establishes the basic ingredients, unitary transport combined with controlled decoherence, upon which genuinely hydrodynamic behaviour could be built. Additional components would be required to produce viscosity, diffusion, or emergent hydrodynamic equations. Nevertheless, the lazy open quantum walk limit explored here provides a clean analytical foundation for such extensions and could form the basis for future quantum fluid algorithms. Assuming continued progress along the UK’s national quantum roadmap, such methods could be deployable on fault-tolerant quantum computers by around 2030 \citep{UKComputeRoadmap2025,UKQuantumStrategy2023}. Recent work by \cite{foulds2025lazyquantumwalksnative} has also presented gate model circuit implementations of the lazy discrete time quantum walk, suggesting compatibility with Rydberg atom architectures that naturally support multiqubit interactions. In order to introduce decoherence into the coin subsystem, the coin qubit(s) can be entangled with one or more ancillary qubits via a controlled-$k$-phase (C$k$Z) gate. The resulting degree of entanglement, tuneable through the state preparation of the ancilla qubit(s), directly controls the effective strength of decoherence acting on the coin. This procedure enlarges the overall Hilbert space and introduces additional junk components into the final state vector, but enables the extraction of the decohered dynamics through suitable post-processing of the ancilla degrees of freedom.

Future research may extend the present framework to higher spatial dimensions, explore wider classes of coin operators, and investigate how breaking or reducing internal coin symmetries can be exploited to model specific physical transport processes. More broadly, the continuum limits derived here provide a systematic route for connecting open quantum walks to macroscopic transport equations, and may serve as a useful starting point for applications in kinetic and hydrodynamic settings. This perspective is particularly relevant in the context of the lattice Boltzmann method, where formal connections between quantum walks and the lattice Boltzmann method have previously been established \citep{Succi_LBM_is_QW}. The continuum limit obtained in the present work yields a linear quantum field theory featuring coherent advection together with simple dephasing induced dissipation. While this captures the leading transport behaviour, more elaborate quantum walk constructions, such as multi-step schemes or additional internal states, are known to generate richer structures in the continuum limit, including dispersive terms \citep{Jolly_2023}. Such models may therefore offer a pathway toward more faithful quantum analogues of lattice kinetic theories. In addition, reformulations based on the Madelung transformation provide an alternative route to introducing nonlinear advection at the level of hydrodynamic variables, without modifying the underlying linear quantum evolution \citep{hatifi2017quantumwalkhydrodynamics}.

\ack{We gratefully acknowledge Professor Giuseppe Di Molfetta for his guidance and many discussions throughout this research. We also thank Dr Steph Foulds for their careful reading of the manuscript and their many insightful comments and suggestions. For the purpose of open access, the authors have applied a Creative Commons Attribution (CC BY) licence to any Author Accepted Manuscript version arising from this submission.}

\funding{VK funded by UKRI EPSRC grants EP/T026715/2, EP/T001062/1, EP/W00772X/2, and EP/Y004566/1, EP/Z53318X/1, and the UKRI DRI and STFC funded CCP-QC Bridge Project -- extended case studies for neutral atom hardware. LJ supported by the UK Engineering and Physical Sciences Research Council (EPSRC) through doctoral funding.}

\appendix

\section{Gell Mann Matrices}\label{app}

Explicitly, the Gell Mann matrices are given by \cite{gellmannmatrices}:
\[
\begin{aligned}
\lambda_1 &= 
\begin{pmatrix} 0 & 1 & 0 \\ 1 & 0 & 0 \\ 0 & 0 & 0 \end{pmatrix}, &
\lambda_2 &= 
\begin{pmatrix} 0 & -i & 0 \\ i & 0 & 0 \\ 0 & 0 & 0 \end{pmatrix}, &
\lambda_3 &= 
\begin{pmatrix} 1 & 0 & 0 \\ 0 & -1 & 0 \\ 0 & 0 & 0 \end{pmatrix},\\[4pt]
\lambda_4 &= 
\begin{pmatrix} 0 & 0 & 1 \\ 0 & 0 & 0 \\ 1 & 0 & 0 \end{pmatrix}, &
\lambda_5 &= 
\begin{pmatrix} 0 & 0 & -i \\ 0 & 0 & 0 \\ i & 0 & 0 \end{pmatrix}, &
\lambda_6 &= 
\begin{pmatrix} 0 & 0 & 0 \\ 0 & 0 & 1 \\ 0 & 1 & 0 \end{pmatrix},\\[4pt]
\lambda_7 &= 
\begin{pmatrix} 0 & 0 & 0 \\ 0 & 0 & -i \\ 0 & i & 0 \end{pmatrix}, &
\lambda_8 &= 
\frac{1}{\sqrt{3}}
\begin{pmatrix} 1 & 0 & 0 \\ 0 & 1 & 0 \\ 0 & 0 & -2 \end{pmatrix}
\end{aligned}
\]


\addcontentsline{toc}{section}{References}

\begin{thebibliography}{49}
\providecommand{\natexlab}[1]{#1}
\providecommand{\url}[1]{\texttt{#1}}
\expandafter\ifx\csname urlstyle\endcsname\relax
  \providecommand{\doi}[1]{doi: #1}\else
  \providecommand{\doi}{doi: \begingroup \urlstyle{rm}\Url}\fi

\bibitem[Aharonov et~al.(1993)Aharonov, Davidovich, and Zagury]{PhysRevA.48.1687}
Y.~Aharonov, L.~Davidovich, and N.~Zagury.
\newblock Quantum random walks.
\newblock \emph{Phys. Rev. A}, 48:\penalty0 1687--1690, Aug 1993.
\newblock \doi{10.1103/PhysRevA.48.1687}.
\newblock URL \url{https://link.aps.org/doi/10.1103/PhysRevA.48.1687}.

\bibitem[Alberti et~al.(2014)Alberti, Alt, Werner, and Meschede]{Alberti_2014}
Andrea Alberti, Wolfgang Alt, Reinhard Werner, and Dieter Meschede.
\newblock Decoherence models for discrete-time quantum walks and their application to neutral atom experiments.
\newblock \emph{New Journal of Physics}, 16\penalty0 (12):\penalty0 123052, December 2014.
\newblock ISSN 1367-2630.
\newblock \doi{10.1088/1367-2630/16/12/123052}.
\newblock URL \url{http://dx.doi.org/10.1088/1367-2630/16/12/123052}.

\bibitem[Ambainis(2004)]{ambainis2004}
Andris Ambainis.
\newblock Quantum walks and their algorithmic applications, 2004.
\newblock URL \url{https://arxiv.org/abs/quant-ph/0403120}.
\newblock arXiv preprint.

\bibitem[Ambainis(2007)]{doi:10.1137/S0097539705447311}
Andris Ambainis.
\newblock Quantum walk algorithm for element distinctness.
\newblock \emph{SIAM Journal on Computing}, 37\penalty0 (1):\penalty0 210--239, 2007.
\newblock \doi{10.1137/S0097539705447311}.
\newblock URL \url{https://doi.org/10.1137/S0097539705447311}.

\bibitem[Ambainis et~al.(2001)Ambainis, Bach, Nayak, Vishwanath, and Watrous]{ambainis2001}
Andris Ambainis, Eric Bach, Ashwin Nayak, Ashvin Vishwanath, and John Watrous.
\newblock One-dimensional quantum walks.
\newblock In \emph{Proceedings of the Thirty-Third Annual ACM Symposium on Theory of Computing}, STOC '01, page 37–49, New York, NY, USA, 2001. Association for Computing Machinery.
\newblock ISBN 1581133499.
\newblock \doi{10.1145/380752.380757}.
\newblock URL \url{https://doi.org/10.1145/380752.380757}.

\bibitem[Annabestani et~al.(2010)Annabestani, Akhtarshenas, and Abolhassani]{deco_in_1d}
Mostafa Annabestani, Seyed~Javad Akhtarshenas, and Mohamad~Reza Abolhassani.
\newblock Decoherence in a one-dimensional quantum walk.
\newblock \emph{Phys. Rev. A}, 81:\penalty0 032321, Mar 2010.
\newblock \doi{10.1103/PhysRevA.81.032321}.
\newblock URL \url{https://link.aps.org/doi/10.1103/PhysRevA.81.032321}.

\bibitem[Attal et~al.(2012{\natexlab{a}})Attal, Petruccione, Sabot, and Sinayskiy]{Attal_2012}
S.~Attal, F.~Petruccione, C.~Sabot, and I.~Sinayskiy.
\newblock Open quantum random walks.
\newblock \emph{Journal of Statistical Physics}, 147\penalty0 (4):\penalty0 832–852, May 2012{\natexlab{a}}.
\newblock ISSN 1572-9613.
\newblock \doi{10.1007/s10955-012-0491-0}.
\newblock URL \url{http://dx.doi.org/10.1007/s10955-012-0491-0}.

\bibitem[Attal et~al.(2012{\natexlab{b}})Attal, Petruccione, and Sinayskiy]{ATTAL20121545}
S.~Attal, F.~Petruccione, and I.~Sinayskiy.
\newblock Open quantum walks on graphs.
\newblock \emph{Physics Letters A}, 376\penalty0 (18):\penalty0 1545--1548, 2012{\natexlab{b}}.
\newblock ISSN 0375-9601.
\newblock \doi{https://doi.org/10.1016/j.physleta.2012.03.040}.
\newblock URL \url{https://www.sciencedirect.com/science/article/pii/S0375960112003453}.

\bibitem[Au-Yeung et~al.(2025)Au-Yeung, Kendon, and Lind]{Au_Yeung_2025}
R.~Au-Yeung, V.~M. Kendon, and S.~J. Lind.
\newblock Quantum smoothed particle hydrodynamics algorithm inspired by quantum walks.
\newblock \emph{Physics of Fluids}, 37\penalty0 (5), May 2025.
\newblock ISSN 1089-7666.
\newblock \doi{10.1063/5.0268240}.
\newblock URL \url{http://dx.doi.org/10.1063/5.0268240}.

\bibitem[Beige et~al.(2000)Beige, Braun, Tregenna, and Knight]{Almut_dissipation}
Almut Beige, Daniel Braun, Ben Tregenna, and Peter~L. Knight.
\newblock Quantum computing using dissipation to remain in a decoherence-free subspace.
\newblock \emph{Phys. Rev. Lett.}, 85:\penalty0 1762--1765, Aug 2000.
\newblock \doi{10.1103/PhysRevLett.85.1762}.
\newblock URL \url{https://link.aps.org/doi/10.1103/PhysRevLett.85.1762}.

\bibitem[Bialynicki-Birula(1996)]{bialynickibirula2005photonwavefunction}
Iwo Bialynicki-Birula.
\newblock V photon wave function.
\newblock volume~36 of \emph{Progress in Optics}, pages 245--294. Elsevier, 1996.
\newblock \doi{https://doi.org/10.1016/S0079-6638(08)70316-0}.
\newblock URL \url{https://www.sciencedirect.com/science/article/pii/S0079663808703160}.

\bibitem[Breuer and Petruccione(2002)]{BreuerPetruccione2002}
Heinz-Peter Breuer and Francesco Petruccione.
\newblock \emph{The Theory of Open Quantum Systems}.
\newblock Oxford University Press, 2002.
\newblock ISBN 9780199213900.

\bibitem[Brun et~al.(2003)Brun, Carteret, and Ambainis]{Brun_Ambainis}
Todd~A. Brun, H.~A. Carteret, and Andris Ambainis.
\newblock Quantum random walks with decoherent coins.
\newblock \emph{Phys. Rev. A}, 67:\penalty0 032304, Mar 2003.
\newblock \doi{10.1103/PhysRevA.67.032304}.
\newblock URL \url{https://link.aps.org/doi/10.1103/PhysRevA.67.032304}.

\bibitem[Childs(2009)]{Childs_2009}
Andrew~M. Childs.
\newblock On the relationship between continuous- and discrete-time quantum walk.
\newblock \emph{Communications in Mathematical Physics}, 294\penalty0 (2):\penalty0 581–603, October 2009.
\newblock ISSN 1432-0916.
\newblock \doi{10.1007/s00220-009-0930-1}.
\newblock URL \url{http://dx.doi.org/10.1007/s00220-009-0930-1}.

\bibitem[di~Molfetta and Debbasch(2012)]{Molfetta_Debbasch_limits_and_sym}
G.~di~Molfetta and F.~Debbasch.
\newblock Discrete-time quantum walks: Continuous limit and symmetries.
\newblock \emph{Journal of Mathematical Physics}, 53\penalty0 (12):\penalty0 123302, 11 2012.
\newblock ISSN 0022-2488.
\newblock \doi{10.1063/1.4764876}.
\newblock URL \url{https://doi.org/10.1063/1.4764876}.

\bibitem[{D\"ur} et~al.(2002){D\"ur}, Raussendorf, Kendon, and Briegel]{Dur_2002}
W.~{D\"ur}, R.~Raussendorf, V.~M. Kendon, and H.-J. Briegel.
\newblock Quantum random walks in optical lattices.
\newblock \emph{Phys.~Rev.~A}, 66:\penalty0 052319, 2002.
\newblock URL \url{https://arxiv.org/quant-ph/abs/0207137}.

\bibitem[Farhi and Gutmann(1998)]{Farhi_1998}
Edward Farhi and Sam Gutmann.
\newblock Quantum computation and decision trees.
\newblock \emph{Physical Review A}, 58\penalty0 (2):\penalty0 915–928, August 1998.
\newblock ISSN 1094-1622.
\newblock \doi{10.1103/physreva.58.915}.
\newblock URL \url{http://dx.doi.org/10.1103/PhysRevA.58.915}.

\bibitem[Foulds and Kendon(2025)]{foulds2025lazyquantumwalksnative}
Steph Foulds and Viv Kendon.
\newblock Lazy quantum walks with native multiqubit gates, 2025.
\newblock URL \url{https://arxiv.org/abs/2511.21608}.
\newblock arXiv preprint.

\bibitem[Giri and Korepin(2019)]{Giri_2019}
Pulak~Ranjan Giri and Vladimir Korepin.
\newblock Lackadaisical quantum walk for spatial search.
\newblock \emph{Modern Physics Letters A}, 35\penalty0 (08):\penalty0 2050043, December 2019.
\newblock ISSN 1793-6632.
\newblock \doi{10.1142/s0217732320500431}.
\newblock URL \url{http://dx.doi.org/10.1142/S0217732320500431}.

\bibitem[Hatifi et~al.(2019)Hatifi, Di~Molfetta, Debbasch, and Brachet]{hatifi2017quantumwalkhydrodynamics}
Mohamed Hatifi, Giuseppe Di~Molfetta, Fabrice Debbasch, and Marc Brachet.
\newblock Quantum walk hydrodynamics.
\newblock \emph{Scientific Reports}, 9:\penalty0 2989, 2019.
\newblock \doi{10.1038/s41598-019-40059-x}.
\newblock URL \url{https://www.nature.com/articles/s41598-019-40059-x}.

\bibitem[Itani and Succi(2021)]{itani2021analysiscarlemannlinearizationlattice}
Wael Itani and Sauro Succi.
\newblock Analysis of {C}arlemann linearization of lattice {B}oltzmann, 2021.
\newblock URL \url{https://arxiv.org/abs/2111.11327}.
\newblock arXiv preprint.

\bibitem[Jolly and Di~Molfetta(2023)]{Jolly_2023}
Nicolas Jolly and Giuseppe Di~Molfetta.
\newblock Twisted quantum walks, generalised {D}irac equation and {F}ermion doubling.
\newblock \emph{The European Physical Journal D}, 77\penalty0 (5), May 2023.
\newblock ISSN 1434-6079.
\newblock \doi{10.1140/epjd/s10053-023-00659-9}.
\newblock URL \url{http://dx.doi.org/10.1140/epjd/s10053-023-00659-9}.

\bibitem[Kemp et~al.(2020)Kemp, Sinayskiy, and Petruccione]{Kemp2020}
Garreth Kemp, Ilya Sinayskiy, and Francesco Petruccione.
\newblock Lazy open quantum walks.
\newblock \emph{Phys. Rev. A}, 102:\penalty0 012220, Jul 2020.
\newblock \doi{10.1103/PhysRevA.102.012220}.
\newblock URL \url{https://link.aps.org/doi/10.1103/PhysRevA.102.012220}.

\bibitem[Kendon(2007)]{KENDON_2007}
Viv Kendon.
\newblock Decoherence in quantum walks – a review.
\newblock \emph{Mathematical Structures in Computer Science}, 17\penalty0 (6):\penalty0 1169–1220, December 2007.
\newblock ISSN 1469-8072.
\newblock \doi{10.1017/s0960129507006354}.
\newblock URL \url{http://dx.doi.org/10.1017/S0960129507006354}.

\bibitem[Kendon(2020)]{kendon_quantum_computation}
Viv Kendon.
\newblock How to compute using quantum walks.
\newblock \emph{Electronic Proceedings in Theoretical Computer Science}, 315:\penalty0 1--17, 04 2020.
\newblock \doi{10.4204/EPTCS.315.1}.

\bibitem[Kendon and Tregenna(2003)]{Kendon-useful}
Viv Kendon and Ben Tregenna.
\newblock Decoherence can be useful in quantum walks.
\newblock \emph{Phys. Rev. A}, 67:\penalty0 042315, Apr 2003.
\newblock \doi{10.1103/PhysRevA.67.042315}.
\newblock URL \url{https://link.aps.org/doi/10.1103/PhysRevA.67.042315}.

\bibitem[L\'opez and Paz(2003)]{Paz_Lopez_deco}
Cecilia~C. L\'opez and Juan~Pablo Paz.
\newblock Phase-space approach to the study of decoherence in quantum walks.
\newblock \emph{Phys. Rev. A}, 68:\penalty0 052305, Nov 2003.
\newblock \doi{10.1103/PhysRevA.68.052305}.
\newblock URL \url{https://link.aps.org/doi/10.1103/PhysRevA.68.052305}.

\bibitem[Love and Boghosian(2005)]{Love2005FromDT}
Peter~J. Love and Bruce~M. Boghosian.
\newblock From {D}irac to diffusion: Decoherence in quantum lattice gases.
\newblock \emph{Quantum Information Processing}, 4:\penalty0 335--354, 2005.
\newblock URL \url{https://api.semanticscholar.org/CorpusID:1461684}.

\bibitem[Manighalam and Kon(2020)]{Manighalam2020}
Michael Manighalam and Mark Kon.
\newblock General methods and properties to evaluate continuum limits of the {1D} discrete time quantum walk.
\newblock \emph{Quantum Information Processing}, 19\penalty0 (10):\penalty0 379, October 2020.
\newblock ISSN 1573-1332.
\newblock \doi{10.1007/s11128-020-02880-6}.
\newblock URL \url{https://doi.org/10.1007/s11128-020-02880-6}.

\bibitem[Mohseni et~al.(2008)Mohseni, Rebentrost, Lloyd, and Aspuru-Guzik]{Mohseni2008}
Masoud Mohseni, Patrick Rebentrost, Seth Lloyd, and Alán Aspuru-Guzik.
\newblock Environment-assisted quantum walks in photosynthetic energy transfer.
\newblock \emph{The Journal of Chemical Physics}, 129\penalty0 (17):\penalty0 174106, 2008.
\newblock \doi{10.1063/1.3002335}.
\newblock URL \url{https://doi.org/10.1063/1.3002335}.

\bibitem[Molfetta(2021)]{dimolfetta2021}
Giuseppe~Di Molfetta.
\newblock Quantum walks, limits and transport equations, 2021.
\newblock URL \url{https://arxiv.org/abs/2112.11828}.
\newblock arXiv preprint.

\bibitem[Molfetta and Arrighi(2019)]{DiMolfetta2019}
Giuseppe~Di Molfetta and Pablo Arrighi.
\newblock A quantum walk with both a continuous-time limit and a continuous-spacetime limit.
\newblock \emph{Quantum Information Processing}, 19\penalty0 (2):\penalty0 47, February 2019.
\newblock ISSN 1573-1332.
\newblock \doi{10.1007/s11128-019-2549-2}.
\newblock URL \url{https://doi.org/10.1007/s11128-019-2549-2}.

\bibitem[Molfetta and Debbasch(2012)]{DiMolfetta2012}
Giuseppe~Di Molfetta and Fran{\c{c}}ois Debbasch.
\newblock Discrete-time quantum walks: Continuous limit and symmetries.
\newblock \emph{Journal of Mathematical Physics}, 53\penalty0 (12):\penalty0 123302, 2012.
\newblock ISSN 1089-7658.
\newblock \doi{10.1063/1.4769171}.
\newblock URL \url{https://doi.org/10.1063/1.4769171}.

\bibitem[Romanelli et~al.(2005)Romanelli, Siri, Abal, Auyuanet, and Donangelo]{ROMANELLI2005137}
A.~Romanelli, R.~Siri, G.~Abal, A.~Auyuanet, and R.~Donangelo.
\newblock Decoherence in the quantum walk on the line.
\newblock \emph{Physica A: Statistical Mechanics and its Applications}, 347:\penalty0 137--152, 2005.
\newblock ISSN 0378-4371.
\newblock \doi{https://doi.org/10.1016/j.physa.2004.08.070}.
\newblock URL \url{https://www.sciencedirect.com/science/article/pii/S0378437104011422}.

\bibitem[Saha et~al.(2021)Saha, Mandal, Saha, and Chakrabarti]{Saha2021}
Amit Saha, Sudhindu Mandal, Debasri Saha, and Amlan Chakrabarti.
\newblock One-dimensional lazy quantum walk in ternary system.
\newblock \emph{IEEE Transactions on Quantum Engineering}, PP:\penalty0 1--1, 04 2021.
\newblock \doi{10.1109/TQE.2021.3074707}.
\newblock URL \url{https://doi.org/10.1109/TQE.2021.3074707}.

\bibitem[Shenvi et~al.(2003)Shenvi, Kempe, and Whaley]{Kempe_Shenvi_Whaley}
Neil Shenvi, Julia Kempe, and K.~Birgitta Whaley.
\newblock Quantum random-walk search algorithm.
\newblock \emph{Phys. Rev. A}, 67:\penalty0 052307, May 2003.
\newblock \doi{10.1103/PhysRevA.67.052307}.
\newblock URL \url{https://link.aps.org/doi/10.1103/PhysRevA.67.052307}.

\bibitem[Sinayskiy and Petruccione(2012)]{Sinayskiy_2012}
Ilya Sinayskiy and Francesco Petruccione.
\newblock Properties of open quantum walks on $\mathbb {Z}$.
\newblock \emph{Physica Scripta}, 2012\penalty0 (T151):\penalty0 014077, nov 2012.
\newblock \doi{10.1088/0031-8949/2012/T151/014077}.
\newblock URL \url{https://doi.org/10.1088/0031-8949/2012/T151/014077}.

\bibitem[Sinayskiy and Petruccione(2014)]{Sinayskiy2014}
Ilya Sinayskiy and Francesco Petruccione.
\newblock Efficiency of open quantum walk implementation of dissipative quantum computing algorithms.
\newblock \emph{Quantum Information Processing}, 13\penalty0 (4):\penalty0 805--813, 2014.
\newblock ISSN 1573-1332.
\newblock \doi{10.1007/s11128-013-0692-z}.
\newblock URL \url{https://doi.org/10.1007/s11128-013-0692-z}.

\bibitem[Sinayskiy and Petruccione(2015)]{Sinayskiy2015}
Ilya Sinayskiy and Francesco Petruccione.
\newblock Microscopic derivation of open quantum walks.
\newblock \emph{Phys. Rev. A}, 92:\penalty0 032105, Sep 2015.
\newblock \doi{10.1103/PhysRevA.92.032105}.
\newblock URL \url{https://link.aps.org/doi/10.1103/PhysRevA.92.032105}.

\bibitem[Succi(2018)]{Succi2018}
Sauro Succi.
\newblock \emph{The Lattice Boltzmann Equation for Complex States of Flowing Matter}.
\newblock Oxford University Press, Oxford, 2018.
\newblock ISBN 9780199592357.
\newblock \doi{10.1093/oso/9780199592357.001.0001}.
\newblock URL \url{https://doi.org/10.1093/oso/9780199592357.001.0001}.

\bibitem[{Succi} et~al.(2015){Succi}, {Fillion-Gourdeau}, and {Palpacelli}]{Succi_LBM_is_QW}
Sauro {Succi}, Fran{\c{c}}ois {Fillion-Gourdeau}, and Silvia {Palpacelli}.
\newblock {Quantum lattice Boltzmann is a quantum walk}.
\newblock \emph{EPJ Quantum Technology}, 2\penalty0 (1):\penalty0 12, December 2015.
\newblock \doi{10.1140/epjqt/s40507-015-0025-1}.
\newblock URL \url{https://ui.adsabs.harvard.edu/abs/2015EPJQT...2...12S}.

\bibitem[Tude and de~Oliveira(2022)]{Tude_2022b}
Luísa~Toledo Tude and Marcos~César de~Oliveira.
\newblock Decoherence in the three-state quantum walk.
\newblock \emph{Physica A: Statistical Mechanics and its Applications}, 605:\penalty0 128012, November 2022.
\newblock ISSN 0378-4371.
\newblock \doi{10.1016/j.physa.2022.128012}.
\newblock URL \url{http://dx.doi.org/10.1016/j.physa.2022.128012}.

\bibitem[{UK Department for Science, Innovation and Technology}(2023)]{UKQuantumStrategy2023}
{UK Department for Science, Innovation and Technology}.
\newblock National {Q}uantum {S}trategy.
\newblock \url{https://www.gov.uk/government/publications/national-quantum-strategy/national-quantum-strategy-accessible-webpage}, 2023.
\newblock Sets out the UK's 10-year plan for quantum technologies, including targets for large-scale, fault-tolerant quantum computers by the early 2030s.

\bibitem[{UK Department for Science, Innovation and Technology}(2025)]{UKComputeRoadmap2025}
{UK Department for Science, Innovation and Technology}.
\newblock {UK} {C}ompute {R}oadmap.
\newblock \url{https://www.gov.uk/government/publications/uk-compute-roadmap/uk-compute-roadmap}, 2025.
\newblock Outlines investment of up to £2 billion through 2030 for advanced and quantum computing infrastructure.

\bibitem[Verstraete et~al.(2009)Verstraete, Wolf, and Cirac]{Verstraete2009}
Frank Verstraete, Michael~M. Wolf, and J.~Ignacio Cirac.
\newblock Quantum computation and quantum-state engineering driven by dissipation.
\newblock \emph{Nature Physics}, 5\penalty0 (9):\penalty0 633--636, 2009.
\newblock \doi{10.1038/nphys1342}.
\newblock URL \url{https://doi.org/10.1038/nphys1342}.

\bibitem[Wilson(2024)]{gellmannmatrices}
Robert~A. Wilson.
\newblock A discrete model for {Gell-Mann} matrices, 2024.
\newblock URL \url{https://arxiv.org/abs/2401.13000}.
\newblock arXiv preprint.

\bibitem[Wong(2015)]{Wong_2015}
Thomas~G Wong.
\newblock Grover search with lackadaisical quantum walks.
\newblock \emph{Journal of Physics A: Mathematical and Theoretical}, 48\penalty0 (43):\penalty0 435304, oct 2015.
\newblock \doi{10.1088/1751-8113/48/43/435304}.
\newblock URL \url{https://doi.org/10.1088/1751-8113/48/43/435304}.

\bibitem[Wong(2018)]{Wong_2018}
Thomas~G. Wong.
\newblock Faster search by lackadaisical quantum walk.
\newblock \emph{Quantum Information Processing}, 17\penalty0 (3), February 2018.
\newblock ISSN 1573-1332.
\newblock \doi{10.1007/s11128-018-1840-y}.
\newblock URL \url{http://dx.doi.org/10.1007/s11128-018-1840-y}.

\bibitem[Yamagishi et~al.(2023)Yamagishi, Hatano, Imura, and Obuse]{Dirac_walk_limit}
Manami Yamagishi, Naomichi Hatano, Ken-Ichiro Imura, and Hideaki Obuse.
\newblock Proposal of multidimensional quantum walks to explore {D}irac and {S}chr\"odinger systems.
\newblock \emph{Phys. Rev. A}, 107:\penalty0 042206, Apr 2023.
\newblock \doi{10.1103/PhysRevA.107.042206}.
\newblock URL \url{https://link.aps.org/doi/10.1103/PhysRevA.107.042206}.

\end{thebibliography}
\end{document}